\documentclass[%
 reprint,
superscriptaddress,
showkeys,
preprintnumbers,
 amsmath,amssymb,
 bibtex,
 aps,
prl
]{revtex4-2}

\usepackage{xcolor}
\usepackage[normalem]{ulem}

\usepackage{fontawesome5}

\usepackage[export]{adjustbox}
\usepackage[colorlinks=true,breaklinks=true]{hyperref}
\hypersetup{urlcolor=blue,
	    citecolor=black}

\usepackage{graphicx}
\usepackage{dcolumn}
\usepackage{bm}
\usepackage{amsmath, amssymb}
\usepackage{graphicx}

\usepackage{multirow}
\usepackage{float}
\usepackage{physics}

\usepackage{tikz}
\usepackage{tkz-euclide}
\usetikzlibrary{shapes.misc}
\usetikzlibrary{decorations.pathmorphing}	
\tikzset{
    v/.style={decorate, decoration={snake, segment length=3mm, amplitude=0.75mm}, draw},
    f/.style={draw=black, postaction={decorate},
        decoration={markings,mark=at position .6 with {\arrow[very thick]{latex}}}},
    fb/.style={draw=black, postaction={decorate},
        decoration={markings,mark=at position .4 with {\arrowreversed[very thick]{latex}}}},
    fnar/.style={draw=black},
    g/.style={decorate, draw=black,
        decoration={coil,amplitude=3pt, segment length=3.5pt}},
    s/.style={dashed,draw=black, postaction={decorate},
        decoration={markings,mark=at position .55 with {\arrow[very thick]{latex}}}},
    sb/.style={dashed,draw=black, postaction={decorate},
        decoration={markings,mark=at position .55 with {\arrowreversed[draw=black,very thick]{latex}}}},
    snar/.style={dashed,draw=black,line width =1.25pt},
    cross/.style={cross out, draw=black, minimum size=2*(#1-\pgflinewidth), inner sep=0pt, outer sep=0pt},
cross/.default={3pt},
}

\def\be{\begin{equation}}
\def\ee{\end{equation}}

\definecolor{palatd}{RGB}{104, 36, 109}
\definecolor{palatb}{RGB}{0, 56, 168}
\definecolor{palatr}{rgb}{0.745,0.118,0.176}
\newcommand\myshade{80}
\colorlet{mylinkcolor}{palatr}
\colorlet{mycitecolor}{palatb}
\colorlet{myurlcolor}{palatd}

\hypersetup{
  linkcolor  = mylinkcolor!\myshade!black,
  citecolor  = mycitecolor!\myshade!black,
  urlcolor   = myurlcolor!\myshade!black,
  colorlinks = true
}

\begin{document}

\preprint{IFT-UAM/CSIC-25-57}

\title{Could a Primordial Black Hole Explosion Explain the extremely high-energy KM3NeT neutrino Event?}

\author{Lua F. T. Airoldi}
\email{lua.airoldi@usp.br}
\affiliation{Instituto de F\'isica, Universidade de S\~ao Paulo, C.P. 66.318, 05315-970 S\~ao Paulo, Brazil}

\author{Gustavo F. S. Alves}%
\email{gustavo.figueiredo.alves@usp.br}
\affiliation{Instituto de F\'isica, Universidade de S\~ao Paulo, C.P. 66.318, 05315-970 S\~ao Paulo, Brazil}

\author{Yuber F. Perez-Gonzalez}%
\email{yuber.perez@uam.es}
\affiliation{Departamento de Física Teórica and Instituto de Física Teórica UAM/CSIC, Universidad Autónoma de Madrid, Cantoblanco, 28049 Madrid, Spain}

\author{Gabriel M. Salla}%
\email{gabriel.massoni.salla@usp.br}
\affiliation{Instituto de F\'isica, Universidade de S\~ao Paulo, C.P. 66.318, 05315-970 S\~ao Paulo, Brazil}

\author{Renata Zukanovich Funchal}%
\email{zukanov@if.usp.br}
\affiliation{Instituto de F\'isica, Universidade de S\~ao Paulo, C.P. 66.318, 05315-970 S\~ao Paulo, Brazil}

\date{\today}

\begin{abstract}
    A black hole is expected to end its lifetime in a cataclysmic runaway burst of Hawking radiation, emitting all Standard Model particles with ultra-high energies. 
    Thus, the explosion of a nearby primordial black hole (PBH) has been proposed as a possible explanation for the $\sim 220$~PeV neutrino-like event recently reported by the KM3NeT collaboration. 
    If the event originated from a PBH, the source would need to lie at $(1\!-\!7)\times10^{-5}\,\mathrm{pc}$—depending on the assumed effective area—thus within the Solar System.
    At such proximity, the resulting flux of gamma rays and cosmic rays would be detectable at Earth. 
    By incorporating the time-dependent field of view of gamma-ray observatories, we show that LHAASO should have recorded ${\cal O}(10^8)$ events between fourteen and seven hours prior to the KM3NeT detection. 
    IceCube and KM3NeT \textit{itself} should likewise have detected of order a few hundred events in the range $1~\mathrm{TeV}\!\lesssim\!E_\nu\!\lesssim\!1~\mathrm{PeV}$ during the 24 hours preceding the burst.
    The absence of any such multi-messenger signal, particularly in gamma-ray data, strongly disfavors the interpretation of the KM3-230213A event as arising from evaporation in a minimal four-dimensional Schwarzschild scenario. 
\end{abstract}

\maketitle


\emph{Introduction. ---} 
The detection of a ${\cal{O}}(100~\rm PeV)$ neutrino event by the KM3NeT/ARCA facility~\cite{KM3NeT:2025npi} 
has sparked a surge of theoretical investigations, each striving to unravel the underlying mechanisms and astrophysical origins of this signal.
Providing a consistent explanation requires addressing 
a critical question: why have other observatories ---most notably the IceCube detector~\cite{IceCube:2016zyt}, having a volume 10 times larger and operated for 15 times longer than KM3NeT--- failed to register similar ultra-high-energy occurrences?
Specifically, if the KM3NeT event, KM3-230213A, arose from an isotropic diffuse flux, it would challenge existing IceCube~\cite{IceCube:2018fhm,IceCube:2024fxo,IceCube:2025ezc} and Pierre Auger~\cite{PierreAuger:2019azx, Savina:2025bir} limits, with recent studies finding a $\sim 2.5~\sigma$–$3.5~\sigma$ tension depending on different assumptions~\cite{KM3NeT:2025ccp,Neronov:2025jfj,Li:2025tqf,Cermenati:2025ogl,Yuan:2025isv}.
Likewise, a long-term steady source is unlikely, as a decade of continuous observations by IceCube has not revealed an emitter at the corresponding event declination~\cite{IceCube:2019cia}.
However, if the KM3NeT event originated from a transient source, the apparent tension with IceCube data diminishes significantly, easing to a more modest discrepancy at the 
$\sim 2 \sigma$ level~\cite{Li:2025tqf}. Recent discussions about KM3-230213A and possible explanations for its origin can be found in Refs.~\cite{Anchordoqui:2025xug, Dev:2025czz, Farzan:2025ydi, Baker:2025cff, Zhang:2025rqh, He:2025bex, Khan:2025gxs, Murase:2025uwv, Das:2025vqd, Zhang:2025abk, Barman:2025hoz, Choi:2025hqt, Dvali:2025ktz, Jho:2025gaf, Klipfel:2025jql, Jiang:2025blz, Crnogorcevic:2025vou, Alves:2025xul, Wang:2025lgn, Narita:2025udw, Kohri:2025bsn, Brdar:2025azm, Borah:2025igh, Boccia:2025hpm, Yang:2025kfr, Neronov:2025jfj, Dzhatdoev:2025sdi, Satunin:2025uui, Muzio:2025gbr}.

A pivotal factor in unraveling the origin of KM3-230213A lies in the conspicuous absence of accompanying gamma rays or other high-energy particles~\cite{Das:2025vqd, Crnogorcevic:2025vou, Wang:2025lgn, Filipovic:2025ulm, Narita:2025udw, Neronov:2025jfj, Dzhatdoev:2025sdi, Fang:2025nzg, Muzio:2025gbr}.
As a result, any proposed novel source capable of generating the ultra-high-energy (UHE) neutrino observed by KM3NeT must either effectively suppress the production of high-energy photons or originate from regions sufficiently distant to attenuate the photon flux during its propagation to Earth due to interactions with intergalactic matter or radiation fields~\cite{Fang:2025nzg}. 

Nevertheless, it has been recently suggested that the evaporation of an {\em asteroid-mass} nearby primordial black hole could be responsible for the KM3NeT neutrino-like event~\cite{Klipfel:2025jql}.
Primordial black holes (PBH), hypothesized to have formed in the early Universe are foreseen to undergo evaporation via Hawking radiation~\cite{Hawking:1974rv} -- a semiclassical effect in which the strong curvature dynamics of collapse generate a  particle flux.
The spectrum of this radiation is thermal~\cite{Hawking:1975vcx}, with a temperature determined by the properties of the PBH. 
For a non-rotating, uncharged black hole, this temperature is inversely proportional to its mass~\cite{Hawking:1975vcx,Page:1976df}.  
As a result of their continuous particle emission, PBHs are predicted to undergo a runaway evaporation process, initially gradual but rapidly accelerating as their mass diminishes. Consequently, only during the terminal phase of this evaporation would particles be emitted with energies comparable to the one observed by KM3NeT on 13 February 2023 at 01:16:47 UTC.

For a PBH to be fully evaporating now its initial mass must be of ${\cal O}(10^{15})$~g, assuming only the presence of Standard Model (SM) degrees of freedom. 
The formation of a black hole with such a low mass necessitates a collapse mechanism fundamentally different from that of stellar origin, which typically yields black holes with masses several times that of the Sun. In contrast, a range of theoretical scenarios has been proposed to account for the formation of these light PBHs, including 
collapse from inhomogeneities, phase transitions, topological defects and bubble collisions~\cite{Carr:2020gox}.

If this scenario could hold, the KM3NeT event would provide the first tantalizing indirect evidence for the existence of PBHs and lend observational support to the long-theorized phenomenon of Hawking radiation.
Additionally, detecting a nearby PBH would support the idea that BHs formed in the early Universe, potentially strengthening their role as a significant Dark Matter component with extended mass distributions, allowing some to evaporate today~\cite{Boluna:2023jlo}.
This is why in this letter we carefully examine 
the feasibility of this hypothesis using a \emph{reductio ad absurdum} approach.

We estimate the distance from Earth a PBH would have to be in order to produce a neutrino flux consistent with the KM3NeT observation to be $\sim (1-7)\times 10^{-5}$ pc.
We show that, in this case, gamma-ray and cosmic-ray observatories should have detected a substantial number of associated photon and proton events in the hours {\em preceding} the final burst.
Specifically, the Large High Altitude Air Shower Observatory (LHAASO)~\cite{LHAASO:2019qtb} would have been sensitive to this emission 
of gamma-rays between 14 and 8 hours before the KM3NeT event, as the corresponding sky region lay within its field of view during that period. 
The final burst would have been within the field of view of the High-Altitude Water Cherenkov (HAWC) gamma-ray observatory~\cite{historical:2023opo}, although HAWC was not operational at that time~\cite{atel17069}. 

Additionally, IceCube~\cite{IceCube:2016zyt} should have observed a rising neutrino flux with energies above $\sim 100~\mathrm{GeV}$ starting about four hours before the final burst, and would have detected $\sim 300$ events at the time of the burst itself.
Moreover, KM3NeT/ARCA itself would have observed $\sim 150$ events in the 11-24 hours prior to the event in the energy range of $1~{\rm TeV}\lesssim E_\nu \lesssim 1~{\rm PeV}$.
Given the absence of any such signals, we conclude that attributing the KM3NeT event to PBH evaporation appears to be highly unlikely and is therefore strongly disfavored as a plausible explanation.

\emph{PBH evaporation. ---} An observer far from a Schwarzschild black hole will detect a flux of thermal radiation for a given particle species $j$, characterized by the instantaneous emission rate per energy bin $\dd E$~\cite{Hawking:1975vcx}
\begin{align}\label{eq:SpectrumEmittedParticles}
    \frac{\dd^2 N_j}{\dd E \dd t} = \frac{g_j}{2\pi} \frac{\Gamma_{s_j}(E)}{\exp(E/T) - (-1)^{2s_j}}\, ,
\end{align}
where $g_j$ and $s_j$ denote the internal degrees of freedom and spin of the particle, respectively~\footnote{Throughout this manuscript, we adopt natural units where $\hbar = c = k_{\rm B} = 1$, and fix the Planck Mass to be $M_P = G^{-1/2}$, with $G$ the gravitational constant}. 
The black hole temperature is inversely proportional to its mass $M$, $T = (8\pi G M)^{-1} \sim 1~{\rm GeV}~(10^{13}~{\rm g}/M)$.
The spectrum in Eq.~\eqref{eq:SpectrumEmittedParticles} differs from a purely Planckian distribution due to the presence of the black hole’s gravitational potential, which acts as an energy-dependent barrier that partially backscatters some emitted particles. This spin-dependent effect is encoded in the absorption probability $\Gamma_{s_j}(E)$, a transmission coefficient of a wave through the effective potential.

From energy conservation considerations, the black hole mass is expected to decrease at a rate that depends on its initial mass and the number of available degrees of freedom. 
The time evolution of the black hole mass can be estimated by calculating the mass loss due to the emission of each particle species, integrated over phase space and summed over all species~\cite{MacGibbon:1990zk,MacGibbon:1991tj,Cheek:2021odj}
\begin{align}\label{eq:dMdt}
\frac{\dd M}{\dd t} = - \sum_j\int \dd E~E \frac{\dd^2 N_j}{\dd E \dd t} \equiv -\frac{\varepsilon(M)}{G^2 M^2},
\end{align}
where $\varepsilon(M)$ is a function that encodes the contribution of all emitted particles at a given mass.
Assuming the emission of only SM degrees of freedom, $\varepsilon(M)\approx 4\times10^{-3}$ for $M \lesssim 10^{10}~{\rm g}$, and decreasing for larger masses.
The mass loss rate in Eq.~\eqref{eq:dMdt} implies that the black hole initially evaporates very slowly, but the process accelerates over time, leading to a runaway evolution that culminates in a final explosive phase. 
Furthermore, as the black hole's temperature increases with time, the energies of the emitted particles also grow, potentially reaching values far beyond those observed to date.

There are, however, important caveats to the discussion of the PBH mass loss rate. First, the derivation of the Hawking spectrum neglects backreaction effects on the spacetime. While this remains a challenging problem, studies in simplified scenarios suggest that, within the semi-classical approximation where fields are treated quantum mechanically and the spacetime classically, the mass loss rate follows $\dd M/\dd t \propto -M^{-2}$~\footnote{Recently, it has been proposed that black hole evaporation may deviate from the standard semi-classical picture due to the so-called ``memory burden'' hypothesis, which suggests modifications to the mass loss rate at late times~\cite{Dvali:2024hsb}. 
This idea is motivated by a possible analogy between certain solitonic solutions in non-Abelian gauge theories and black holes. 
Within this framework, the emission of UHE neutrinos capable of producing the KM3NeT event has also been considered~\cite{Boccia:2025hpm,Dvali:2025ktz,Zantedeschi:2024ram}.
As the hypothesis remains speculative, we assume the semi-classical approximation remains valid up to near-Planckian scales.}. 
We will therefore adopt Eq.~\eqref{eq:dMdt} as our working assumption.
This approximation is expected to break down as the black hole mass approaches the Planck scale $M_P$, where the \emph{adiabaticity condition}, $\dot{T}/T^2 \ll 1$, fails~\cite{Barcelo:2010pj}. Although this regime is not relevant for our purposes, we conservatively stop the evolution at $M \approx 100\,M_P$.
A further complication is the information loss problem stated in terms of entropies: the von Neumann entropy of Hawking radiation eventually exceeds the Bekenstein-Hawking entropy of the black hole after the \emph{Page time}~\cite{Page:1993wv,Page:2013dx,Perez-Gonzalez:2025try}. Beyond this point, non-thermal corrections to the Hawking spectrum may arise from unknown physics. As the black holes considered here should have passed the Page time long ago, such effects could in principle modify the emitted spectrum~\cite{Perez-Gonzalez:2025try}. However, due to the current lack of consensus on their nature or necessity~\cite{Buoninfante:2021ijy,Buoninfante:2025gqk}, we will continue to assume that the emission remains thermal, as given by Eq.~\eqref{eq:SpectrumEmittedParticles}, and that mass loss follows Eq.~\eqref{eq:dMdt}.

\begin{figure}[t]
    \centering
    \includegraphics[width=\linewidth]{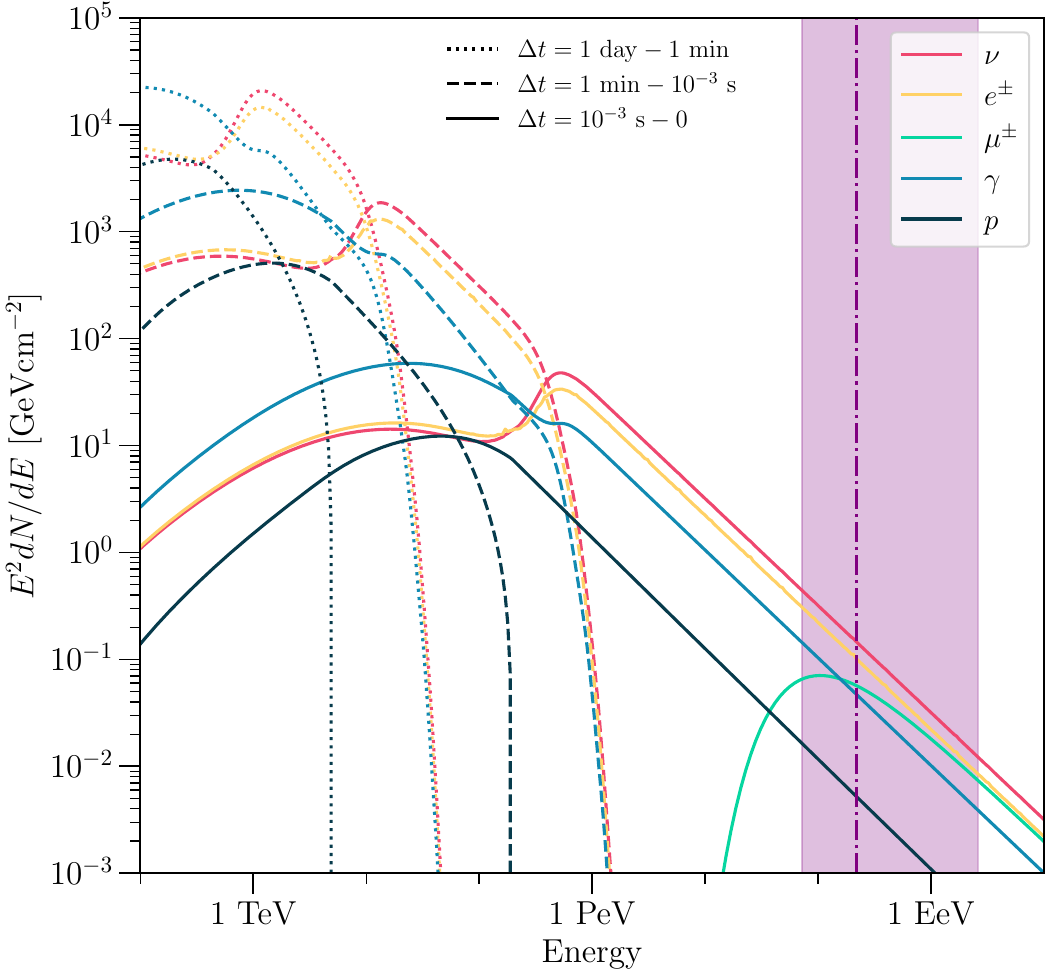}
    \caption{Time integrated Hawking spectrum times energy-squared for three different time intervals: $\Delta t =$ 1 day to 1 minute (dotted lines), 1 minute to $10^{-3}$ ms (dashed lines), and $10^{-3}$ ms to 0 s (solid lines) before complete evaporation, assuming a PBH located at a distance of $2.4\times 10^{-5}$ pc from Earth. We consider the spectra of neutrinos (red), electrons and positrons (yellow), muons (green), photons (blue), and protons (dark blue). The purple band indicates the energy range associated to the KM3-230213A event at 90\% CL.}
    \label{fig:fluence-times}
\end{figure}
In addition to the direct emission of particles, referred to as \emph{primary emission}, the subsequent decay of unstable particles, as well as the hadronization of quarks and gluons emitted by the black hole, generates an additional flux of particles such as $\nu$, $\gamma$, $e^{\pm}$, $\mu^\pm$, and $p$. This secondary component enhances the overall particle spectrum. 
We have used {\tt BlackHawk}~\cite{Arbey:2019mbc,Arbey:2021mbl} to compute the secondary spectra, specifically using the tables obtained using {\tt HDMSpectra}~\cite{Bauer:2020jay}.
Consequently, if a PBH evaporates sufficiently close to Earth, we expect not only a flux of high-energy neutrinos but also other significantly boosted particles.
To illustrate this effect, Fig.~\ref{fig:fluence-times} presents the time-integrated Hawking spectra for three distinct time intervals before the final evaporation: $\Delta t =$ 1 day to 1 minute (dotted), 1 minute to $10^{-3}$ ms (dashed), and $10^{-3}$ ms to 0 (solid), assuming a PBH located at a distance of $2\times 10^{-5}$ pc from Earth. The spectra are shown for all neutrino species (red), electrons and positrons (yellow), muons (green), photons (blue), and protons (dark blue).
For muons, we account for their possible decay during propagation to Earth. As a result, their contribution becomes significant only during the final stages of evaporation, since only then they are produced with very high energies, $E_{\mu^\pm} \gtrsim 10~{\rm PeV}$, which may allow them to reach Earth before decaying.
Additionally, the proton flux, arising from the decay and hadronization of highly boosted, strongly interacting particles, also exhibits high energies, although its magnitude is approximately one order of magnitude lower than that of the other components.

Since the detection of charged particles and photons is governed by electromagnetic interactions, the number of events expected from a PBH evaporation, assuming it is sufficiently close, should be several orders of magnitude higher than the corresponding neutrino events. This consideration might lead to the naive expectation that attributing the KM3-230213A event to a PBH evaporation is in strong tension with the absence of signals in other observatories.
However, there are important subtleties that must be taken into account before drawing such a conclusion. 

\begin{figure*}[t!]
  \centering
  \includegraphics[width=0.8\textwidth]{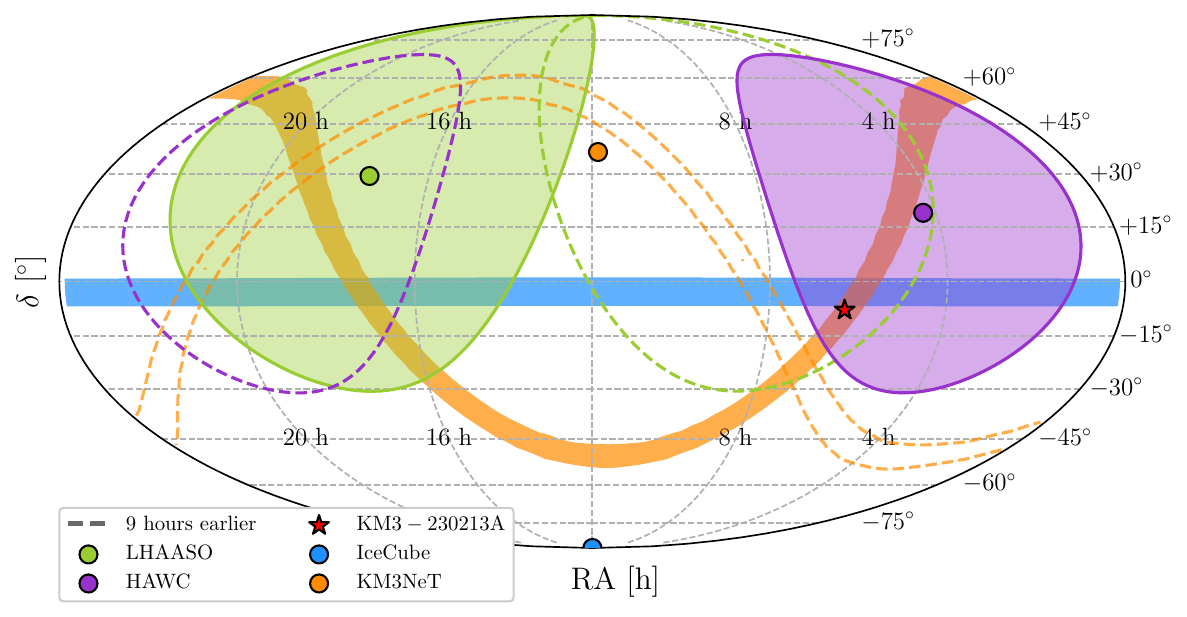}
    \caption{Field of view in equatorial coordinates covered by gamma-ray experiments HAWC~\cite{historical:2023opo,HAWC:2019wla} (purple) and LHAASO~\cite{DiSciascio:2016rgi,LHAASO:2019qtb} (light green), together with the regionswhere the effective areas of KM3NeT (orange) and IceCube (light blue) are $25\%$ below their maximum value for an energy equal to the measured by KM3NeT. 
    The colored regions denote the instantaneous field of view of these experiments at the time of the KM3NeT event, 13th of February 2023 at 01:16:47 UTC~\cite{KM3NeT:2025npi} (red star). 
    The area encompassed by the dashed lines represent the corresponding regions nine hours earlier.}
    \label{fig:coverage}
\end{figure*}
\emph{KM3-230213A as an evaporating PBH. ---} Let us assume that the KM3NeT event originated from a neutrino produced by the evaporation of a nearby PBH. Under this hypothesis, our first objective is to determine the distance to Earth, $d_\text{PBH}$, at which the PBH evaporated. 
Following IceCube~\cite{Dave:2019epr}, the total number of neutrino-induced events expected in KM3NeT is given by
\begin{align}\label{eq:Nevts}
N_\text{evt}^\nu = \frac{1}{4\pi d_\text{PBH}^2} \int_0^\tau dt \int \dd E  \frac{\dd^2 N_{\nu}}{\dd E \dd t} A_\text{eff}(E,\delta_\text{PBH},\text{RA}_\text{PBH}),
\end{align}
where $\frac{\dd^2 N_{\nu}}{\dd E \dd t}$ includes both primary and secondary neutrino contributions, $\tau$ is an initial remaining evaporation lifetime, and $A_\text{eff}(E,\delta_\text{PBH},\text{RA}_\text{PBH})$ denotes the detector’s effective area, depending on neutrino energy and the PBH celestial coordinates, specified by its declination $\delta_\text{PBH}$ and right ascension $\text{RA}_\text{PBH}$.
Vacuum neutrino oscillations are incorporated in the determination of number of events, as described in detail in the companion paper~\cite{Airoldi:2025bgr}.

It is important to emphasize that, unlike searches for diffuse neutrino fluxes in observatories such as IceCube or KM3NeT, the dependence on the event’s right ascension must be explicitly taken into account. This is because, for transient sources with durations shorter than one hour, the instantaneous orientation of the detector relative to the source plays a crucial role in determining the possible arrival directions of high-energy neutrinos, thereby significantly affecting the effective area.
We use the publicly available code from the Planetary Neutrino Monitoring network (PLE$\nu$M)~\cite{Schumacher:2021hhm,Schumacher:2025qca}~\href{https://github.com/PLEnuM-group/Plenum}{\faGithubSquare} to determine the instantaneous effective areas. The procedure to obtain $A_\text{eff}(E,\delta_\text{PBH},\text{RA}_\text{PBH})$ is provided in detail in Ref.~\cite{Airoldi:2025bgr}. 
For the KM3NeT case, we scale the PLE$\nu$M effective area by 0.2 to account for the reduced ARCA volume at the time of the event~\cite{Muzio:2025gbr}. As a cross-check, we also use the zenith-binned $A_{\rm eff}$ from the full detector configuration~\cite{KM3NeT:2024paj}, rescaled by the same factor, as well as the all-sky–averaged $A_{\rm eff}$ reported in Ref.~\cite{KM3NeT:2025npi}.

Discarding, for the moment, contributions from particles other than neutrinos, and adopting the equatorial coordinates of the evaporating PBH as reported by the KM3NeT collaboration, we should integrate over a time interval corresponding to the experiment’s livetime of $\tau=335$ days. 
However, since KM3NeT/ARCA is optimized for neutrinos above $\sim$~TeV, only the final day of PBH evaporation contributes significantly, see Fig.~\ref{fig:fluence-times}. To estimate $d_{\rm PBH}$, we integrate $N_{\rm evt}^\nu$ over $E_\nu \in [1~{\rm PeV}, 1~{\rm EeV}]$ to cover the relevant energy range of the event and set $N_{\rm evt}^\nu = 1$.
Using the rescaled PLE$\nu$M effective area, we obtain $d_\text{PBH} = 2.4\times 10^{-5}$~pc, while the all-sky–averaged and rescaled zenith-binned $A_{\rm eff}$ yield distances of $0.8 \times 10^{-5}$ and $6.6 \times 10^{-5}$~pc, respectively~\footnote{Certainly, the PBH could be located at a larger distance such that the KM3NeT event would result from a Poissonian upward fluctuation. However, assuming an upward-fluctuation with a probability of 1\%, the PBH distance would be enlarged by a factor of $\sim 10$, which does not affect our conclusions.}.
At such proximity, a substantial flux of other particles, including gamma-rays, UHE electrons, positrons, protons, and potentially muons, would also be expected.

Assuming that the PBH emits only SM degrees of freedom, we compute the time-integrated spectra of photons and protons, as well as the expected number of events in observatories such as LHAASO~\cite{LHAASO:2019qtb}, HAWC~\cite{historical:2023opo}. 
These gamma-ray facilities, which are sensitive to photons with energies above $\sim 100$~GeV, would be capable of detecting such a burst. 
Furthermore, given the assumed proximity of the evaporating PBH, attenuation of the gamma-ray flux during propagation is not expected to be significant. 
Although other instruments, such as Fermi~\cite{Thompson:2022ufx} and the Pierre Auger Observatory~\footnote{We estimate the Pierre Auger Observatory~\cite{PierreAuger:2015eyc} could have observed about 3 gamma-ray 
events with $E_\gamma >5 \times 10^{9}$ GeV during the final burst.}, could also be sensitive to the resulting UHE fluxes, we consider that focusing on the high-energy gamma-ray observatories mentioned above is sufficient to test this hypothesis.

A key consideration in estimating the number of detectable events at these observatories is their limited field of view (FOV). For example, LHAASO has an instantaneous FOV of approximately $60^\circ$, which restricts its sky coverage at any given time. Consequently, it is essential to determine which portion of the sky each observatory was monitoring during the period leading up to the PBH evaporation.
This is illustrated in Fig.~\ref{fig:coverage}, where we show the FOVs of LHAASO and HAWC at the time of the KM3NeT event. For comparison, we also indicate the regions of the sky where the effective areas of IceCube and KM3NeT are at most $25\%$ lower than their maximum values, for neutrinos with energies equal to that of the KM3-230213A event.
We find that, at the time of the final evaporation, which is set to coincide with the KM3NeT event, the corresponding region of the sky was not within the field of view of LHAASO. However, it did lie within the field of view of HAWC at that time, which was not operational at that time~\cite{atel17069}.

\begin{figure}[t!]
    \centering
    \includegraphics[width=\linewidth]{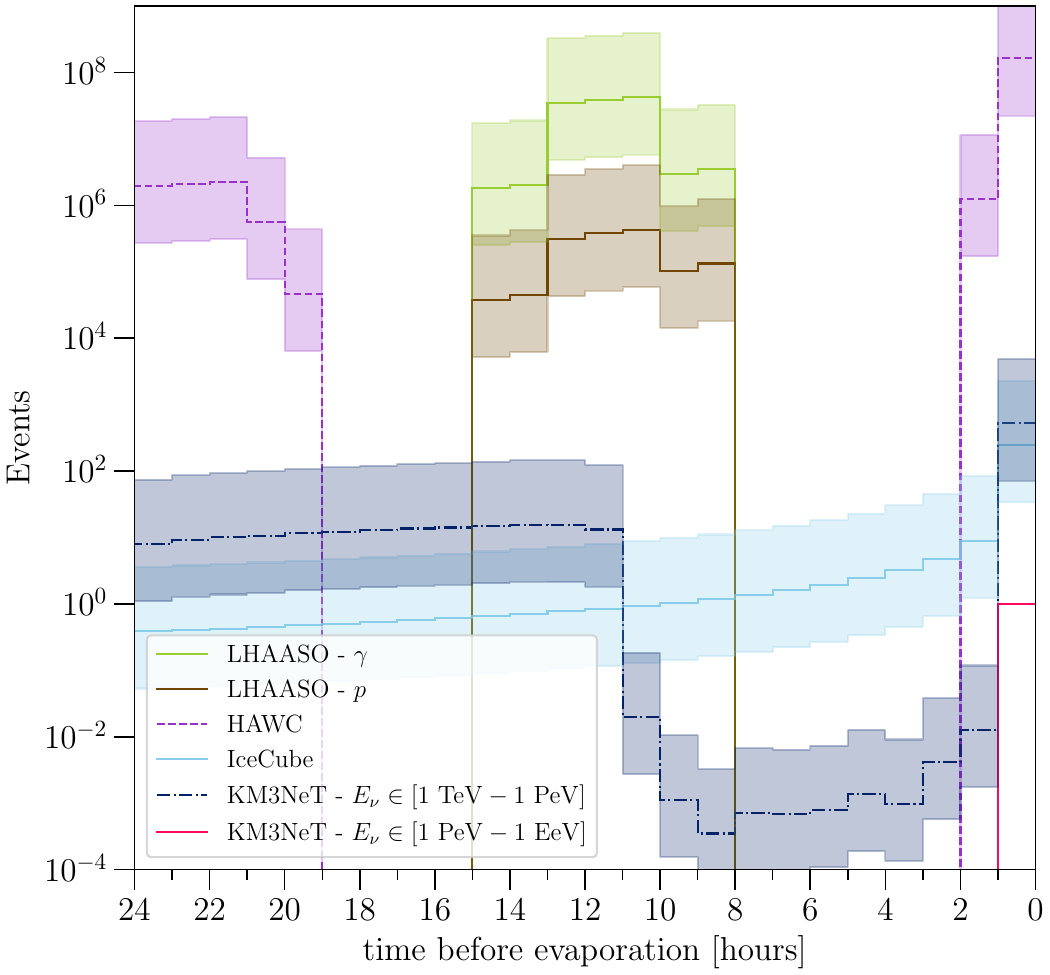}
    \caption{Expected number of events at Earth during the final day before the evaporation of a PBH located at the KM3-230213A sky position, assuming a distance of $2.4\times10^{-5}$ pc. Shown are photon (green) and proton (brown) counts in LHAASO, HAWC photons (lilac dashed), and neutrino events in IceCube (light blue) and KM3NeT for $E_\nu \in [1~\mathrm{TeV}, 1~\mathrm{PeV}]$ (dark blue) and $E_\nu \in [1~\mathrm{PeV}, 1~\mathrm{EeV}]$ (red). Bands reflect events obtained when varying the PBH distance from $10^{-5}$ to $7\times10^{-5}$ pc.}
    \label{fig:events}
\end{figure}
This result suggests that the hypothesis identifying KM3-230213A as the signature of an evaporating PBH could still be viable. However, given the significant proximity of the PBH to Earth, the particle flux in the hours preceding the final burst may be substantial enough that gamma-ray observatories could have detected a flux of high-energy photons originating from the direction of KM3-230213A.
For instance, nine hours before the event, the signal would lie within LHAASO's FOV, see Fig.~\ref{fig:coverage}, dashed lines.
To estimate the number of such events, we need to relate the source position at $(\delta, \text{RA})$ to the local zenith angle at a given time. For this, we use the relation
\begin{align}
    \cos{\theta} = \sin{\lambda} \sin{\delta} + \cos{\lambda} \cos{\delta} \cos\left(2\pi \frac{t}{T} - \text{RA}\right),
\label{eq:equatorial_to_local}
\end{align}
where $\lambda$ is the detector's latitude, $t$ is the local sidereal time, and $T$ is the duration of one sidereal day~\cite{PierreAuger:2019azx}.
To determine the expected number of events one day prior to the final burst, we use the effective areas of LHAASO for photons and protons considering the water cherenkov array detector (WCDA), provided in Ref.~\cite{Yang:2024vij}, across different bins of local zenith angle. If the local zenith angle falls outside LHAASO's field of view, we assume the effective area to be zero.

In Fig.~\ref{fig:events}, we show the expected number of events during the final day of a PBH evaporation located at KM3-230213A's coordinates and at a distance of $2.4\times10^{-5}$ pc. Photon (green) and proton (brown) events for LHAASO, photon events for HAWC (lilac dashed), and neutrino events for IceCube (light blue) and KM3NeT in the energy ranges $E_\nu \in [1\,{\rm TeV},1\,{\rm PeV}]$ (dark blue dot-dashed) and $[1\,{\rm PeV},1\,{\rm EeV}]$ (red) are displayed. The colored bands indicate the variation obtained by varying the PBH distance between $10^{-5}$ and $7\times10^{-5}$ pc. 
Although LHAASO would not have observed the final hour, it would nonetheless have detected a very large number of earlier events: roughly $\mathcal{O}(10^8)$ photons and $\mathcal{O}(10^6)$ protons, in the 14–7 hour window preceding evaporation. IceCube would also register a substantial neutrino flux, collecting approximately  $\sim 300$
events during the final moments. KM3NeT itself would record a significant number of events both in the 11–24 hour interval and in the last hour, for neutrino energies $E_\nu \in [1\,{\rm TeV},1\,{\rm PeV}]$, surpassing IceCube’s expectations in those time windows. Despite KM3NeT being considerably smaller in instrumented volume than IceCube, its effective area is enhanced for this scenario because the KM3-230213A direction remains horizontal for IceCube but becomes upgoing for KM3NeT during part of the day, boosting KM3NeT’s sensitivity at $\mathcal{O}({\rm TeV\!-\!PeV})$ energies~\footnote{We note however that KM3NeT effective areas from PLE$\nu$M are obtained from IceCube ones by a coordinate transformation. Thus, implicitly, these assume the same cuts and selections used by IceCube, which might not be applicable for KM3NeT.}.

Given the unlikelihood that such a significant pre-burst signal would go undetected, we conclude that the hypothesis identifying KM3-230213A as the result of an evaporating PBH is strongly disfavored by the non-observation of correlated signals in other messengers.

\emph{Final Thoughts. ---}
The discovery of the evaporation of a nearby PBH would represent a major milestone in fundamental physics, providing direct evidence for Hawking radiation and thus confirming our understanding of quantum field theory in a dynamic curved spacetime. Furthermore, such an observation could offer new insights into longstanding problems, such as the information loss paradox, through the direct measurement of the black hole time evolution.
The UHE neutrino event reported by the KM3NeT collaboration in February 2023 sparked considerable interest, raising the intriguing possibility that it could be the first observational hint of PBH evaporation. However, our analysis demonstrates that this interpretation does not withstand scrutiny when viewed within the broader observational framework. Given that an evaporating PBH is expected to emit the full spectrum of SM particles, a genuine signal would likely manifest across multiple messengers -- not solely in the neutrino channel.
By accounting for the time-dependent field of view of gamma-ray observatories, we have found that the LHAASO experiment should have detected over ${\cal O}(10^8)$ UHE photon events from the same region of the sky, several hours prior to the KM3NeT detection. The absence of such a signal strongly disfavors the PBH evaporation hypothesis.

More generally, our results apply to any source that emits both photons and neutrinos and is sufficiently close to Earth for photon attenuation to be negligible. To this end, we have developed a general framework for identifying transient sources in current and future neutrino telescopes, highlighting the critical role played by the source's equatorial coordinates in determining detectability.
Our findings underscore the importance of multi-messenger observations in both identifying and localizing transient events. A detailed analysis of this methodology and its broader implications will be presented in a companion paper~\cite{Airoldi:2025bgr}.

\begin{acknowledgments}
\emph{Note added.---} During the finalization of this work, Refs.~\cite{Baker:2025cff,Anchordoqui:2025xug} appeared, proposing that the KM3NeT event could be explained by quasi-extremal or higher-dimensional PBH, respectively. While both studies acknowledge the tension with the non-observation of gamma rays, our analysis complements these efforts by quantitatively demonstrating that this tension excludes the minimal scenario of a four-dimensional Schwarzschild PBH evaporating in close proximity to Earth.

\emph{Acknowledgments. ---}
We thank Pedro Machado, Matheus Hostert, 
Lisa Schumacher, Edivaldo Moura Santos, Daniel Naredo, and Miguel A. Sánchez-Conde for discussions.
We also thank the anonymous Referees for useful suggestions, specially regarding the possibility of KM3NeT/ARCA observing events below the PeV range.
G.F.S.A., L.F.T.A and G.M.S. received full financial support from the São Paulo Research Foundation (FAPESP) through the following contracts No. 2022/10894-8 and No. 2020/08096-0, 2025/07427-7,  2020/14713-2 and 2022/07360-1, respectively.
 R.Z.F. is partially supported by FAPESP under contract No. 2019/04837-9, and by  Conselho Nacional de Desenvolvimento Científico e Tecnológico (CNPq).
Y.F.P.G. was supported by the Consolidaci\'on Investigadora grant CNS2023-144536 from the Spanish Ministerio de Ciencia e Innovaci\'on (MCIN) and by the Spanish Research Agency (Agencia Estatal de Investigaci\'on) through the grant IFT Centro de Excelencia Severo Ochoa No CEX2020-001007-S. 
\end{acknowledgments}

\appendix

\vspace{1em}

\bibliographystyle{apsrev4-1}
\bibliography{km3net_event}

\end{document}